\documentclass[%
aps,
pre,
superscriptaddress,
tightenlines,
showpacs,showkeys,
a4paper,
12pt,
longbibliography,
notitlepage
]{revtex4-1}

\usepackage[T2A]{fontenc}
\usepackage[utf8]{inputenc}
\usepackage[english]{babel}
\usepackage{amssymb,amsmath,stmaryrd}

\usepackage{graphicx}
\usepackage{subfig}

\makeatletter

\newcommand{\sca}[2]{\ensuremath{\bigl({#1}\cdot{#2}\bigr)}}
\newcommand{\avr}[1]{\ensuremath{\langle{#1}\rangle}}

\newcommand{\pdrs}[1]{\partial_{#1}}

\newcommand{\bnbl}{\boldsymbol{\nabla}}

\newcommand{\mum}{$\mu$m}
\newcommand{\dega}{$^\circ$}

 \newcommand{\bs}[1]{\boldsymbol{#1}}
 \newcommand{\vc}[1]{\mathbf{#1}}
 \newcommand{\mvc}[1]{\mathbf{#1}}
 \newcommand{\uvc}[1]{\hat{\mathbf{#1}}}
 
 \newcommand{\ind}[1]{\mathrm{#1}}

\newcommand{\dd}{\mathrm{d}}

 \newcommand{\e}{\mathrm{e}}

\makeatother

\begin{document}
\title{%
  Waveguide propagation of light  in polymer porous films
  filled with nematic liquid crystals
}

\author{A.D.~Kiselev}
\email[Email address: ]{alexei.d.kiselev@gmail.com}
\affiliation{%
  Saint Petersburg National Research University of Information
  Technologies,
  Mechanics and Optics (ITMO University),
Kronverksky Prospekt 49, 197101 Saint Petersburg, Russia
}

\author{S.V.~Pasechnik}
\email[Email address: ]{s-p-a-s-m@mail.ru}
\affiliation{%
MIREA - Russian Technological University,
Vernadskogo Ave., 78, Moscow 119454, Russia}

\author{D.V.~Shmeliova}
\email[Email address: ]{shmeliova@mail.ru}
\affiliation{%
MIREA - Russian Technological University,
 Vernadskogo Ave., 78, Moscow 119454, Russia}

\author{A.P.~Chopik}
\affiliation{%
MIREA - Russian Technological University,
 Vernadskogo Ave., 78, Moscow 119454, Russia}

\author{D.A.~Semerenko}
\affiliation{%
MIREA - Russian Technological University,
 Vernadskogo Ave., 78, Moscow 119454, Russia}

\author{A.V.~Dubtsov}
\email[Email address: ]{dav.e@mail.ru}
\affiliation{%
MIREA - Russian Technological University,
 Vernadskogo Ave., 78, Moscow 119454, Russia}

\date{\today}

\begin{abstract}
  We theoretically analyze
the waveguide regime of
light propagation in a cylindrical pore of a polymer matrix filled
with liquid crystals
assuming that the effective radial optical anisotropy
is biaxial.
From numerical analysis of the dispersion relations,
the waveguide modes are found to be sensitive
to the field-induced changes of the anisotropy.
The electro-optic properties of the polymer porous
polyethylene terephthalate
(PET) films filled with the nematic liquid crystal 5CB
are studied experimentally and
the experimental results
are compared with the results of the theoretical investigation. 
\end{abstract}


 \maketitle

\section{Introduction}
\label{sec:intro}

Liquid crystals (LCs) are known to be of primary importance
in the modern display industry.
A unique combination
of physical properties of LCs
provides a means
to control light propagation through
micron-sized LC layers using electric fields
at low operating voltages and small energy consumption.
The key feature of liquid crystal materials
that underlies
excellent electro-optical
characteristics of LCs
and makes them promising for photonic
applications 
is the
high sensitivity of LC orientational structures
to both external fields and
conditions at bounding surfaces.
In particular, the use of LC in microstructured
waveguides~\cite{Corella:optcomm:2006,Tkachenko:jopt:2008}
and
waveguides based on photonic
crystals~\cite{Wahle:optexp:2014,Marcos:rmp:2017}
is mainly determined by
this feature providing an efficient tool
to adjust optical
characteristics of
waveguide photonic devices
by means of the external (electric) field.

Utilization of LCs in nondisplay applications
such as photonic devices~\cite{Beeckman:opteng:2011} and
a variety of sensors ~\cite{Pasechnik:bk:2009,Carlton:lcrev:2013}
faces a number of challenges.
These include a slow response time of orientational
transformations for the devices of THz photonics
based on thick layers of LC~\cite{Pan:mclc:2004},
high operating voltages needed
to tune propagation of light in photonics fibers filled
with LC~\cite{Larsen:optexp:2003}
and relatively low sensitivity of planar LC layers to impurities
in biosensensing applications~\cite{Carlton:lcrev:2013}.
Some of these problems can be solved by usage of composite materials, like
PDLC, LC emulsions and porous films filled with LC. 

In this paper the electro-optical properties of
the porous polymer films filled with nematic liquid
crystals will be our primary concern.
Recently such type of composite material~---~the
polymer polyethylene terephthalate (PET) film~---~with
normally oriented open pores of submicron and micron sizes,
filled with LC was experimentally studied
in~\cite{Semerenko:ol:2010,Chopik:optlett:2014}.
It was shown that such systems
are very promising for applications in
photonic waveguide and terahertz (THz) devices~\cite{Chopik:optlett:2014,Pasechnik:dd:2016}.

In particular,
we have observed pronounced changes of light
intensity transmitted through the film induced by applying
low frequency ac electric field.
Similarly, modulation of the light intensity
can also be caused by heating of the sample via
absorption of blue light by
the azo-dye layer adsorbed on
the internal surfaces of the pores.
The hypothesis put forward in~\cite{Chopik:optlett:2014}
to explain the experimental results
assumes that breaking of
the waveguide regime of light
propagation is responsible for
the light modulation.

These results motivate theoretical
considerations of this
paper aimed to examine the waveguide regime
of light propagation in cylindrical LC pores.
In our calculations,
we shall treat the LC material as an effective medium with
variable anisotropy embedded in an isotropic environment.
Though similar approach was previously
used in~\cite{Tkachenko:jopt:2008},
it was not applied to analyze
the waveguide regime.
Rigorous analysis of the problem based
on numerical solution of Maxwell’s equations~\cite{Lin:ol:1992}
showed that waveguide propagation of light in a cylindrical cavity
surrounded by isotropic media is crucially influenced by
the orientational structure formed inside
the pore~\cite{Reyes:pre:2002,Rodrigues:pre:2003,Corella:physb:2008}.
From the other hand,
the type of the orientational configuration is governed by
a number of parameters such as the Frank elastic constants,
the surface anchoring strengths and orientation of the easy axes,
the diameter of LC pores~\cite{Alld:prl:1991,Craw:prl2:1993,Kralj:lc:1993,Kiselev:jetp:1995,Kiselev:mclc:1995}.
The paper is organized as follows.

In Sec.~\ref{sec:theory}
we present the details of our theoretical analysis
and systematically study the effects of effective LC anisotropy on
the waveguide regime of light propagation.
In Sec.~\ref{sec:experiment},
we describe the experimental results obtained for
the porous PET films with micron and submicron pores
filled with LC 5CB.
In our discussion of these results we, in particular,
demonstrate a qualitative agreement between
the experimental data 
and the theoretical predictions.
Concluding remarks are given in Sec.~\ref{sec:conclusion}.

\section{Theory}
\label{sec:theory}

\subsection{Basic equations}
\label{subsec:basic_eqs}

We begin with the Maxwell equations for a monochromatic electromagnetic wave
\begin{align}
  \label{eq:Maxwell}
  \bnbl\times\vc{E}=ik_0\mu\vc{H},
\quad
 \bnbl\times\vc{H}=-ik_0\vc{D},
\quad
\vc{D}=\boldsymbol{\epsilon}\vc{E},
\end{align}
where
$k_0=\omega/c$ is the free space wavenumber
and
$\bs{\epsilon}$ is the dielectric tensor,
that propagate along the axis of
a cylindrical waveguide (the $z$ axis).
Assuming that the electromagnetic field can be written in
the factorized form
$\{\vc{E},\vc{H}\}\e^{i(k_z z-\omega t)}$,
we shall use the cylindrical coordinate system
$\{\rho, \phi, z\}$ and
separate out the longitudinal
and transverse parts of both the electromagnetic field
and the gradient operator
as follows:
\begin{align}
&
  \label{eq:transverse}
  \vc{E}=\vc{E}_T+E_z \uvc{z},
\quad
\bnbl=\bnbl_T+\uvc{z}\,\pdrs{z}=
\bnbl_T+ik_z\uvc{z},
\\
&
\label{eq:nabla_t}
\bnbl_T=\uvc{x}\,\pdrs{x}+\uvc{y}\,\pdrs{y}=
\uvc{e}_{\rho}\,\pdrs{\rho}+
\uvc{e}_{\phi}\,\rho^{-1}\,\pdrs{\phi},
\end{align}
where
$\uvc{e}_{\rho}=\cos\phi\,\uvc{x}+\sin\phi\,\uvc{y}$
and
$\uvc{e}_{\phi}=-\sin\phi\,\uvc{x}+\cos\phi\,\uvc{y}$.
With the help of the relation
$\uvc{z}\times(\bnbl\times\vc{E})=\bnbl_T E_z-ik_z\vc{E}_T$,
we obtain
the transverse part of Maxwell's equations~\eqref{eq:Maxwell}
in the form:
\begin{align}
  \label{eq:trns-1}
  \bnbl_T E_z-ik_z\vc{E}_T=ik_0\mu\, \uvc{z}\times\vc{H}_T,
\quad
(\uvc{z}\times\bnbl_T) H_z-ik_z\,\uvc{z}\times\vc{H}_T=ik_0\, \vc{D}_T.
\end{align}
In what follows, we, following
the results of the paper~\cite{Tkachenko:jopt:2008},
restrict ourselves to the case of
effective radial biaxial anisotropy
characterized by the constitutive relations
of the form:
\begin{align}
  \label{eq:D_Tz}
  \vc{D}_T=\bs{\epsilon}_T\vc{E}_T=
  \begin{pmatrix}
    \epsilon_{\rho}& 0\\
0&\epsilon_{\phi}
  \end{pmatrix}
   \begin{pmatrix}
     E_\rho\\
E_{\phi}
   \end{pmatrix}
,
\quad
D_z=\epsilon_{zz}E_z.
\end{align}
Note that the case of radial anisotropy was also considered in the
theory of light scattering by optically anisotropic cylindrical
particles developed in Refs.~\cite{Gao:jpier:2010,Chen:pra:2012}. 

From equations~\eqref{eq:trns-1}
it is not difficult to express the transverse components of
the electromagnetic field 
$\{\vc{E}_T,\vc{H}_T\}$,
in terms of
the longitudinal components
$\{E_z, H_z\}$,
as follows:
\begin{align}
  \label{eq:Et-Ez}
  &
i(k_z^2-k_0^2\mu\bs{\epsilon}_T)\vc{E}_T=
k_z\bnbl_T E_z-k_0\mu(\uvc{z}\times\bnbl_T)H_z,
\\
&
\label{eq:Ht-Hz}
i(k_z^2-k_0^2\mu\bs{\epsilon}_T)\uvc{z}\times\vc{H}_T=
k_z(\uvc{z}\times\bnbl_T) H_z-k_0\bs{\epsilon}_T\,\bnbl_T\, E_z.
\end{align}
For convenience,
we shall introduce the dimensionless parameters
\begin{align}
  \label{eq:parameters}
  r=k_0\rho,\quad
q_z=k_z/k_0,\quad
q_\rho^2=\mu\epsilon_\rho-q_z^2=n_\rho^2-q_z^2,
\quad
q_\phi^2=\mu\epsilon_\phi-q_z^2=n_\phi^2-q_z^2
\end{align}
and write down the expressions for
the transverse components in the explicit form:
\begin{align}
&
  \label{eq:E_rho}
  E_{\rho}=\frac{i}{q_\rho^2 r}
\{
q_z r\pdrs{r}E_z+\mu \pdrs{\phi}H_z
\},
\\
&
\label{eq:E_phi}
  E_{\phi}=\frac{i}{q_\phi^2 r}
\{
q_z \pdrs{\phi}E_z
-\mu r\pdrs{r}H_z
\},
\\
&
  \label{eq:H_rho}
  H_{\rho}=\frac{i}{q_\phi^2 r}
\{
q_z r \pdrs{r}H_z-\epsilon_{\phi} \pdrs{\phi}E_z
\},
\\
&
\label{eq:H_phi}
  H_{\phi}=\frac{i}{q_\rho^2 r}
\{
q_z \pdrs{\phi}H_z
+\epsilon_{\rho} r \pdrs{r}E_z
\}.
\end{align}

Equations for the longitudinal components can be
obtained using the divergence relations:
$\sca{\bnbl}{\vc{D}}=\sca{\bnbl}{\vc{H}}=0$
(these follow from Maxwell's equations~\eqref{eq:Maxwell}),
which can be conveniently recast into the form:
\begin{align}
  \label{eq:divD}
  \sca{\bnbl_T}{\vc{D}_T}=-ik_zD_z,
\quad
  \sca{\bnbl_T}{\vc{H}_T}=\sca{\uvc{z}\times\bnbl_T}{\uvc{z}\times\vc{H}_T}=-ik_zH_z.
\end{align}
After eliminating the transverse components from
the relations~\eqref{eq:Et-Ez}--~\eqref{eq:Ht-Hz}
with the help of Eq.~\eqref{eq:divD},
we have
\begin{align}
  \label{eq:Helm-Ez}
&
\Bigl\{
\bnbl_T\bs{\epsilon}_T\,
\bigl[
k_z^2-k_0^2\mu\,\bs{\epsilon}_T
\bigr]^{-1}
\bnbl_T
-\epsilon_{zz}
\Bigr\} 
E_z=
q_z
\bnbl_T\,
\bigl[
k_z^2-k_0^2\mu\,\bs{\epsilon}_T
\bigr]^{-1}
(\uvc{z}\times\bnbl_T)H_z
\\
&
  \label{eq:Helm-Hz}
\Bigl\{
\uvc{z}\times\bnbl_T\,
\mu
\bigl[
k_z^2-k_0^2\mu\,\bs{\epsilon}_T
\bigr]^{-1}
\uvc{z}\times\bnbl_T
-\mu
\Bigr\} 
H_z=
q_z
(\uvc{z}\times\bnbl_T)\,
\bigl[
k_z^2-k_0^2\mu\,\bs{\epsilon}_T
\bigr]^{-1}
\bnbl_T\,
E_z,
\end{align}
where the operators are given by
\begin{align}
&
  \label{eq:nabla:1}
-r^2
  \bnbl_T\bs{\epsilon}_T\,
\bigl[
k_z^2-k_0^2\mu\,\bs{\epsilon}_T
\bigr]^{-1}
\bnbl_T=
\frac{\epsilon_\rho}{q_{\rho}^2}\,
[r\pdrs{r}]^2+
\frac{\epsilon_\phi}{q_{\phi}^2}\,
\pdrs{\phi}^2,
\\
&
\label{eq:nabla:2}
-r^2
 \uvc{z}\times\bnbl_T\,
\bigl[
k_z^2-k_0^2\mu\,\bs{\epsilon}_T
\bigr]^{-1}
\uvc{z}\times\bnbl_T=
\frac{1}{q_{\phi}^2}\,
[r\pdrs{r}]^2+
\frac{1}{q_{\rho}^2}\,
\pdrs{\phi}^2,
\\
&
\label{eq:coupling}
-r^2\bnbl_T\,
\bigl[
k_z^2-k_0^2\mu\,\bs{\epsilon}_T
\bigr]^{-1}
(\uvc{z}\times\bnbl_T)
=
r^2(\uvc{z}\times\bnbl_T)\,
\bigl[
k_z^2-k_0^2\mu\,\bs{\epsilon}_T
\bigr]^{-1}
\bnbl_T=
\notag
\\
&
\Bigl(
\frac{1}{q_\phi^2}
-\frac{1}{q_\rho^2}
\Bigr)
r\pdrs{r}\pdrs{\phi}=
\frac{\mu(\epsilon_\rho-\epsilon_{\phi})}{q_\rho^2q_\phi^2}r\pdrs{r}\pdrs{\phi}.
\end{align}
For the Fourier harmonics of the longitudinal components
\begin{align}
  \label{eq:EH_m}
  E_z=E_m(r)\e^{im\phi},
\quad
  H_z=-iH_m(r)\e^{im\phi},
\quad
\pdrs{\phi}\to im,
\end{align}
where $m\in\mathbb{Z}$
is the azimuthal number
enumerating the harmonics,
we derive the following system of equations:
\begin{align}
  \label{eq:system}
  \begin{pmatrix}
   \epsilon_\rho q_{\phi}^2\,
[r\pdrs{r}]^2-
\epsilon_\phi q_{\rho}^2\,
m^2+\epsilon_{zz} q^2 r^2& m \kappa_z r\pdrs{r}\\
m \kappa_z r\pdrs{r}&
\mu q_{\rho}^2\,
[r\pdrs{r}]^2-
\mu q_{\phi}^2\,
m^2+\mu q^2 r^2
  \end{pmatrix}
                                  \begin{pmatrix}
                                    E_m\\
                                    H_m
                                  \end{pmatrix}
=\vc{0},
\end{align}
where
$q^2=q_{\rho}^2\,q_{\phi}^2$, 
$\kappa_z=q_z\mu\Delta\epsilon$
and
$\Delta\epsilon=\epsilon_\rho-\epsilon_{\phi}$.

\subsection{Boundary conditions and dispersion relations}
\label{sec:boundary-conds}

For the special case of a zero mode with $m=0$,
the solution of the system~\eqref{eq:system}
can be readily found and is given by
\begin{align}
  \label{eq:m0}
  E_z=E_0(r)= A_0 J_0(r_{z\rho}q_\rho r),
\quad
  iH_z=H_0(r)= B_0 J_0(q_\phi r),
\quad
r_{z\rho}^2=\frac{\epsilon_{zz}}{\epsilon_{\rho}}.
\end{align}
In general case,
we introduce the matrix notations
\begin{align}
  \label{eq:A_alp}
  \mvc{A}(\alpha)=
  \begin{pmatrix}
   \epsilon_\rho q_{\phi}^2\,
\alpha^2-
\epsilon_\phi q_{\rho}^2\,
m^2& m \kappa_z \alpha\\
m \kappa_z \alpha&
\mu q_{\rho}^2\,
\alpha^2-
\mu q_{\phi}^2\,
m^2
  \end{pmatrix},
\quad
\mvc{B}=
 q^2 \begin{pmatrix}
 \epsilon_{zz}  & 0\\
0&
\mu 
  \end{pmatrix}
\end{align}
and will search for the solutions of the system~\eqref{eq:system}
that are regular at $r=0$
(on the waveguide axis)
in the form of a power series:
\begin{align}
  \label{eq:series}
  \begin{pmatrix}
    E_{m,\alpha}(r)\\
H_{m,\alpha}(r)
  \end{pmatrix}=
r^{\alpha}\sum_{n=0}^{\infty}\vc{c}_n r^n.
\end{align}
Substitution of the expansion~\eqref{eq:series}
into Eq.~\eqref{eq:system}
gives the equation for
the coefficient
$\vc{c}_0$ and
the recurrence relations for $\vc{c}_n$
\begin{align}
  \label{eq:c_n}
  \mvc{A}_0\vc{c}_0=0,
\quad
\vc{c}_{n}=-\vc{A}_n^{-1}\vc{B}\vc{c}_{n-2},
\quad
\vc{A}_n\equiv\vc{A}(\alpha+n).
\end{align}
The values of the parameter $\alpha$
can be found from the singularity
condition for
the matrix
$\vc{A}_0$
\begin{align}
  \label{eq:alpha}
  \det\vc{A}(\alpha)=
n_\rho^2 q^2 \alpha^{4}-\alpha^2 m^2
  (n_\phi^2q_\rho^4+n_\rho^2q_\phi^4+\kappa_z^2)+
n_\phi^2 q^2 m^{4}=0,
\end{align}
giving two positive values of the parameter
$\alpha\in\{\alpha_{+},\alpha_{-}\}$
that define two solutions which are regular at $r=0$.
Note that,
similar to the case of
Bessel functions of the first kind
$J_\nu(x)$,
we can use the expression for
the determinant of the matrix
$\vc{A}_n$
(at $\alpha=\alpha_{\pm}$):
\begin{align}
  \label{eq:det_An}
  \det\vc{A}_n=
n_\rho^2 q^2 [(\alpha+n)^{4}- \alpha^{4}]
-(m \kappa_z)^2
[(\alpha+n)^{2}- \alpha^{2}]
\end{align}
to prove convergence
of the power series~\eqref{eq:series},
representing an entire function.

The obtained solution can be used not only for analyzing the waveguide
regime, but also for describing
the light scattering by radially anisotropic
scatterers without recourse to
the simplifying assumptions
such as $E_z=0$
used in the literature~\cite{Gao:jpier:2010,Chen:pra:2012}.

\begin{figure*}[!tbh]
\centering
\subfloat[$m=0$ and $l=1$]{
  \resizebox{80mm}{!}{\includegraphics*{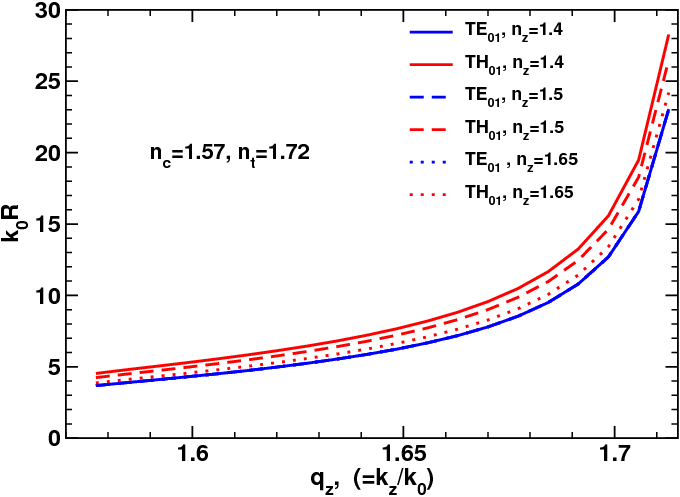}}
\label{subfig:mzero-1}
}
\subfloat[$m=0$ and $l=2$]{
  \resizebox{80mm}{!}{\includegraphics*{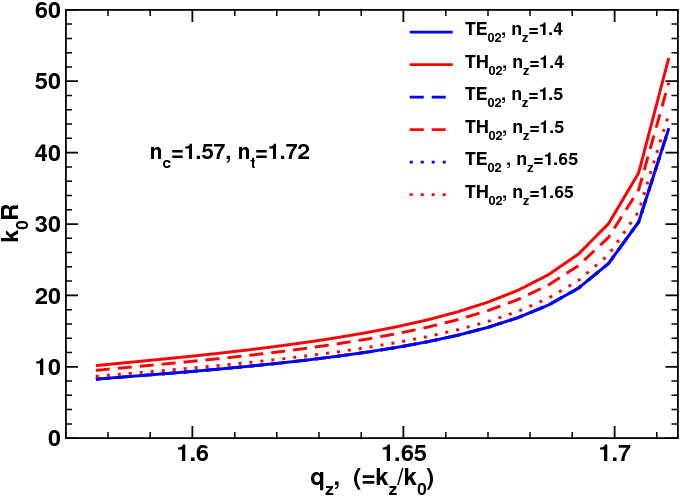}}
\label{subfig:mzero-2}
}
\caption{%
  Dependence of the dimensionless parameter
$k_0 R\equiv \omega R/c$
on $q_z=k_z/k_0$
for 
$\mathrm{TE}_{ml}$
($E_z=0$)
and
$\mathrm{TH}_{ml}$
($H_z=0$)
waves 
with $m=0$
(a)~$l=1$ and (b)~$l=2$
(index $l$ enumerates the branches of roots in ascending order)
at different values of the refractive index $n_z$. 
}
\label{fig:mzero}
\end{figure*}

\begin{figure*}[!tbh]
\centering
\subfloat[$m=0$ and $n_z=1.4$]{
  \resizebox{80mm}{!}{\includegraphics*{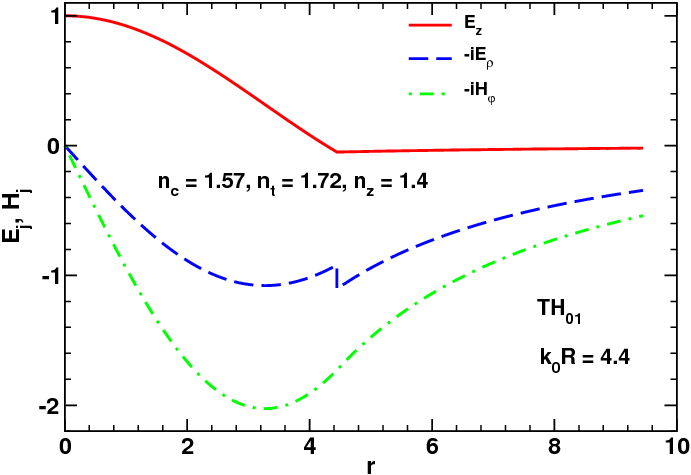}}
\label{subfig:TH-1_4}
}
\subfloat[$m=0$ and $n_z=1.5$]{
  \resizebox{80mm}{!}{\includegraphics*{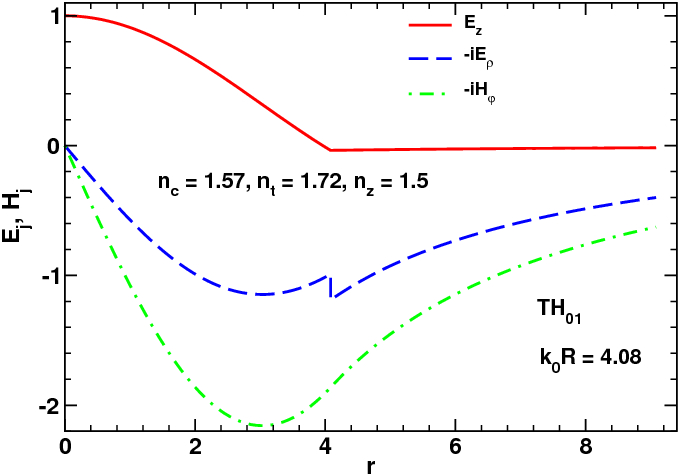}}
\label{subfig:TH-1_5}
}
\caption{%
Electromagnetic field versus
the dimensionless radial parameter 
$r=k_0 \rho$
for the mode
$\mathrm{TH}_{01}$
($H_z=0$)
at
(a)~$n_z=1.4$ and (b)~$n_z=1.5$. 
}
\label{fig:TH}
\end{figure*}

\begin{figure*}[!tbh]
\centering
\subfloat[$m=0$ and $n_z=1.4$]{
  \resizebox{80mm}{!}{\includegraphics*{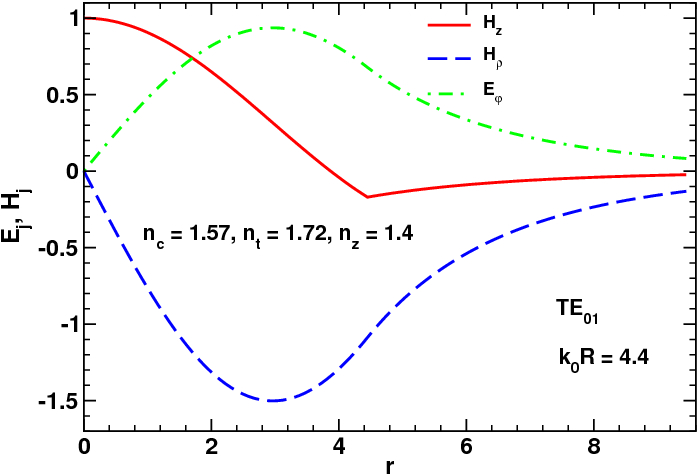}}
\label{subfig:TE-1_4}
}
\subfloat[$m=0$ and $n_z=1.5$]{
  \resizebox{80mm}{!}{\includegraphics*{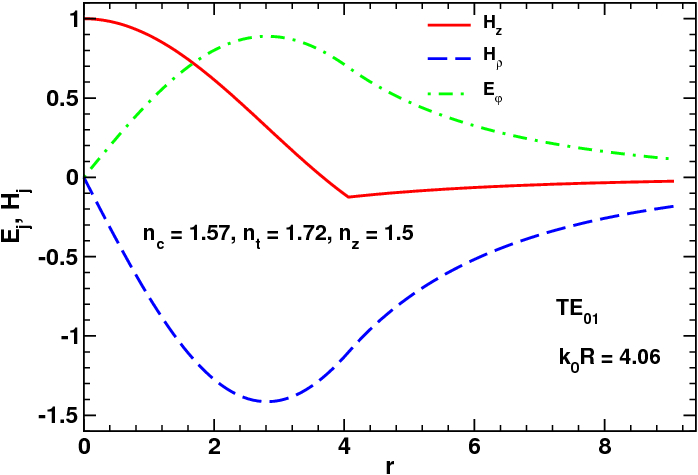}}
\label{subfig:TE-1_5}
}
\caption{%
  Electromagnetic field versus
the dimensionless radial parameter
$r=k_0 \rho$
for the mode
$\mathrm{TE}_{01}$
($E_z=0$)
at
(a)~$n_z=1.4$ and (b)~$n_z=1.5$. 
}
\label{fig:TE}
\end{figure*}

Now,
following Ref.~\cite{Tkachenko:jopt:2008},
we concentrate on the important
special case with
$\epsilon_{\rho}=\epsilon_\phi\equiv\epsilon_t$.
In this case we have
\begin{align}
  \label{eq:EH_m_uni}
  E_z=E_m(r)= A_m J_m(r_{zt}q_t r),
\quad
  iH_z=H_m(r)= B_m J_m(q_t r),
\quad
r_{zt}^2=\frac{\epsilon_{zz}}{\epsilon_{t}},
\end{align}
where
$q_t^2=n_t^2-q_z^2$.

Outside the waveguide at
$r>k_0R\equiv x$,
the solution that exponentially decreases
at infinity,
$E_z, H_z\propto \e^{-r}/\sqrt{r}$,
is given by
\begin{align}
  \label{eq:EH_c_uni}
  E_z=E_m(r)= a_m K_m(q_c r),
\quad
  iH_z=H_m(r)= b_m K_m(q_c r),
\quad
q_c^2=q_z^2-n_c^2
\end{align}
where $K_m(x)$ is the modified Bessel function~\cite{Abr}
and
$n_c=\sqrt{\mu_c\epsilon_c}$ is the refractive index of the
ambient medium.

\begin{figure*}[!tbh]
\centering
\subfloat[$m=1$]{
  \resizebox{80mm}{!}{\includegraphics*{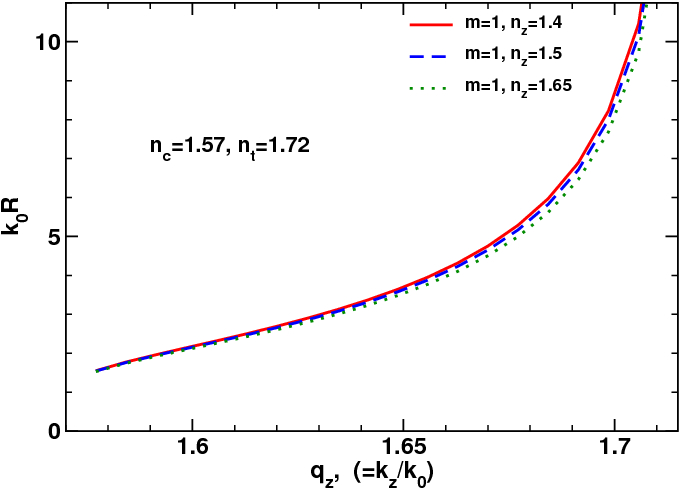}}
\label{subfig:m1}
}
\subfloat[$m=2$]{
  \resizebox{80mm}{!}{\includegraphics*{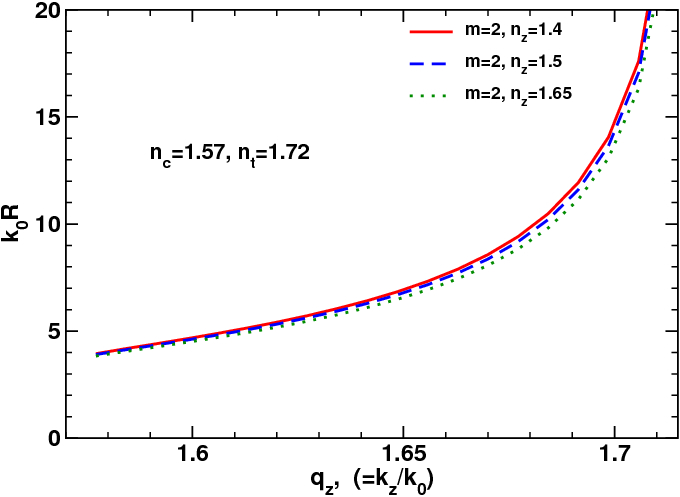}}
\label{subfig:m2}
}
\caption{%
  Dependence of the dimensionless parameter
$k_0 R\equiv \omega R/c$
on $q_z=k_z/k_0$
for the waves with 
(a)~$m=1$ and (b)~$m=2$
at different values of the refractive index
$n_z$. 
}
\label{fig:m12}
\end{figure*}

From the continuity conditions for
the tangential components of the electromagnetic field
at the waveguide boundary
\begin{align}
&
  \label{eq:bc-z}
  E_z(x+0)=E_z(x-0),\quad
  H_z(x+0)=H_z(x-0),
\\
&
\label{eq:bc-phi}
  E_\phi(x+0)=E_\phi(x-0),\quad
  H_\phi(x+0)=H_\phi(x-0),
\quad
x\equiv k_0 R
\end{align}
we have
\begin{align}
&
  \label{eq:alp-bet_m}
\mvc{C}_m
  \begin{pmatrix}
    \alpha_m\\
\beta_m
  \end{pmatrix}
=\mvc{0},
\quad
  \alpha_m =A_mJ_m(r_{zt}x_t)=a_mK_m(x_c),
\quad
  \beta_m =B_mJ_m(x_t)=b_mK_m(x_c),
\end{align}
where $x_t=q_t x$,
$x_c=q_c x$
and the matrix $\vc{C}_m$
is given by
\begin{align}
&
  \label{eq:C_m}
\mvc{C}_m=
  \begin{pmatrix}
    q_z m (n_t^2-n_c^2) & 
\mu q_c^2 R_J^{(m)}(x_t)+\mu_c q_t^2 R_K^{(m)}(x_c)\\
\epsilon_t q_c^2 R_J^{(m)}(r_{zt}x_t)+\epsilon_c q_t^2 R_K^{(m)}(x_c)&q_z m (n_t^2-n_c^2)
  \end{pmatrix},
\\
&
\label{eq:R_JK}
R_J^{(m)}(x)=\frac{xJ_m'(x)}{J_m(x)}=m-\frac{x J_{m+1}(x)}{J_{m}(x)},
\quad
R_K^{(m)}(x)=\frac{xK_m'(x)}{K_m(x)}=m-\frac{x K_{m+1}(x)}{K_{m}(x)}.
\end{align}
Dispersion relations
for the harmonics of the electromagnetic
field,
linking
$k_z$ and $k_0=\omega/c$,
can be deduced from the singularity
condition for the matrix~\eqref{eq:C_m}:
\begin{align}
  \label{eq:disp-rel}
  \det\vc{C}_m(q_z,x)=0,
\quad
n_c^2\le q_z^2=(k_z/k_0)^2\le n_t^2,
\quad
x=k_0 R.
\end{align}

\begin{figure*}[!tbh]
\centering
\resizebox{90mm}{!}{\includegraphics*{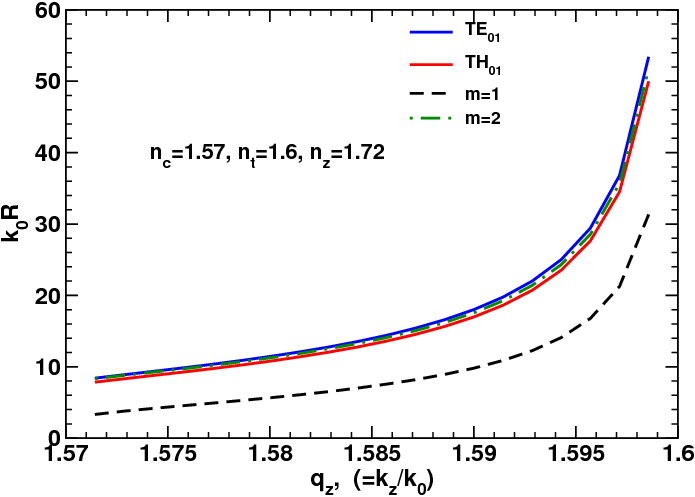}}
\caption{%
Dimensionless parameter
$k_0 R\equiv \omega R/c$
as a function of $q_z=k_z/k_0$
for the waves with
$0\le m\le 2$ 
at
$n_z=1.72>n_t=1.6\approx n_c$.
}
\label{fig:nz-172}
\end{figure*}

\subsection{Numerical results}
\label{subsec:results}

We begin our analysis with the case of waveguide modes with
the zero azimuthal number: $m=0$.
In this case, the dispersion relations
are simplified and reduced to
the relations for two types of the waves:
$TE$ waves with $E_z=0$
\begin{align}
&
  \label{eq:TE}
\mu q_c^2 R_J^{(0)}(x_t)=-\mu_c q_t^2 R_K^{(0)}(x_c)
\Rightarrow
\frac{\mu J_1(x_t)}{x_tJ_0(x_t)}=-\frac{\mu_c K_1(x_c)}{x_cK_0(x_c)}
\end{align}
and
$TH$ waves with $H_z=0$
\begin{align}
  \label{eq:TH}
\epsilon_t q_c^2 R_J^{(0)}(r_{zt}x_t)=-\epsilon_c q_t^2 R_K^{(0)}(x_c)
\Rightarrow
\frac{\epsilon_z J_1(r_{zt}x_t)}{r_{zt}x_tJ_0(r_{zt}x_t)}=-\frac{\epsilon_c K_1(x_c)}{x_cK_0(x_c)}.  
\end{align}
The roots of Eqs.~\eqref{eq:TE} and~\eqref{eq:TH}
are located between the corresponding zeros of
the Bessel function $J_0$
and will be numbered in ascending order
by the index $l$.
In our subsequent calculations,
we shall use the estimates
taken from the paper~\cite{Chopik:optlett:2014}:
the pores of radius
$R$
ranged from
$200$~nm to $600$~nm
are filled with the nematic liquid crystal 5CB
characterized by the refractive indices
$n_e\approx 1.72$ and $n_o\approx 1.5$
($\mu=1$),
the effective refractive index of the
polymer matrix is
$n_c\approx 1.57$
($\mu_c=1$),
the wavelength of the laser radiation
is $\lambda=632$~nm.
Clearly, in this case,
we have
$2\approx 1.98 \le k_0R\le 5.96\approx 6$.

\begin{figure*}[!tbh]
\centering
\subfloat[$m=1$ and $n_z=1.5$]{
  \resizebox{80mm}{!}{\includegraphics*{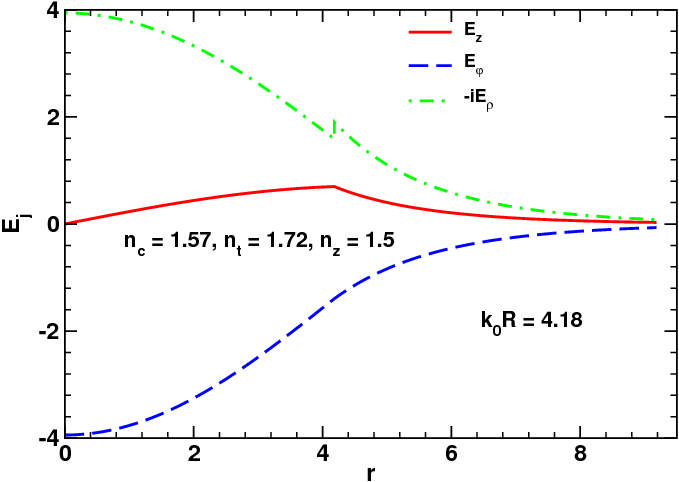}}
\label{subfig:HybE-1_5}
}
\subfloat[$m=1$ and $n_z=1.72$]{
  \resizebox{80mm}{!}{\includegraphics*{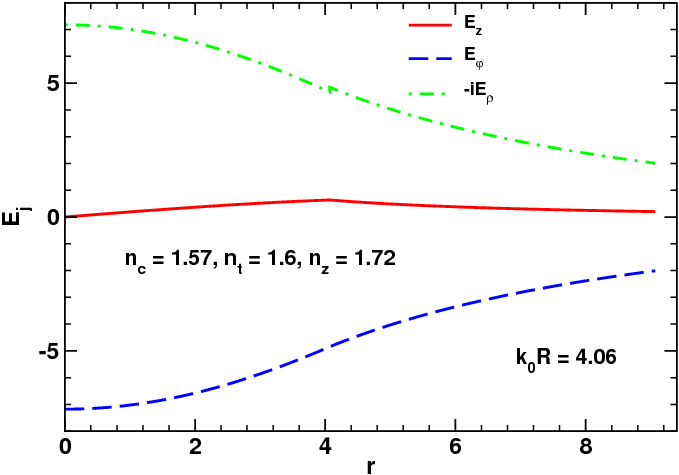}}
\label{subfig:HybE-1_72}
}
\caption{%
  Components of electromagnetic
  filed as a function of the dimensionless
  radial parameter
$r=k_0 \rho$
for the mode
with $m=1$
at
(a)~$n_z=1.5$ and (b)~$n_z=1.72$. 
}
\label{fig:HybE}
\end{figure*}

\begin{figure*}[!tbh]
\centering
\subfloat[$m=1$ and $n_z=1.5$]{
  \resizebox{80mm}{!}{\includegraphics*{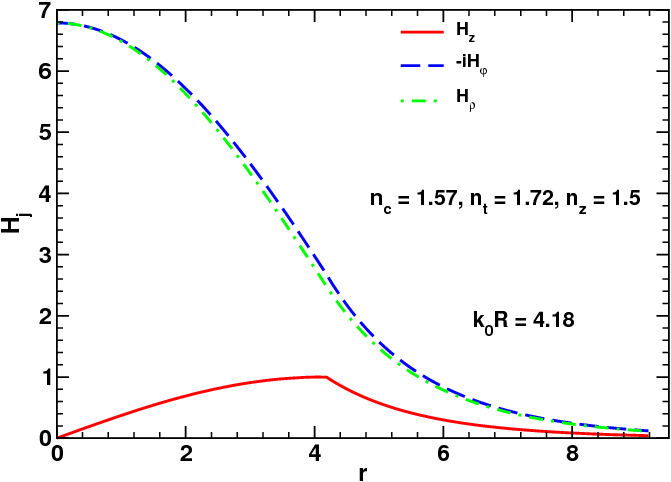}}
\label{subfig:HybH-1_5}
}
\subfloat[$m=1$ and $n_z=1.72$]{
  \resizebox{80mm}{!}{\includegraphics*{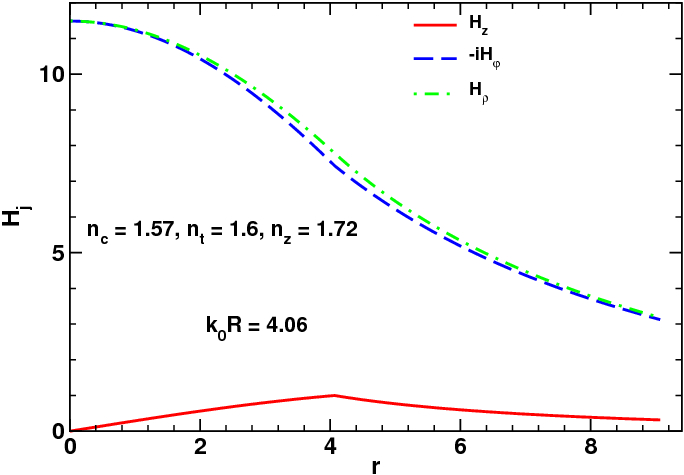}}
\label{subfig:HybH-1_72}
}
\caption{%
  Components of electromagnetic
  filed as a function of the dimensionless
  radial parameter
$r=k_0 \rho$
for the mode
with $m=1$
at
(a)~$n_z=1.5$ and (b)~$n_z=1.72$. 
}
\label{fig:HybH}
\end{figure*}

Figure~\ref{fig:mzero}
shows how
liquid crystal anisotropy 
( the value of the longitudinal refractive index $n_z$)
affect
the dispersion curves for the
waveguide mode
$TE_{0l}$ and $TH_{0l}$.
It can be seen that the curves for the
$TE$ modes,
which are computed for the planar director configuration
with the effective refractive index 
$n_t=n_e=1.72$,
are independent of $n_z$
(see Eq.~\eqref{eq:TE}),
whereas the difference between
the dispersion curves for
the $TH$ and $TE$ modes
becomes more pronounced
as the difference between
the refractive indices
$n_t$ and $n_z$ increases.
The radial distributions
of nonvanishing components
of the electromagnetic fields
for the waveguide modes
$TH_{01}$ and $TE_{01}$
are presented in Figs.~\ref{fig:TH}
and~\ref{fig:TE}, respectively.

The results illustrating
effect of
$n_z$ on the hybrid modes 
($E_z\ne 0$ and $H_z\ne 0$) with 
$m=1$ and $m=2$
are plotted in Fig.~\ref{fig:m12}.
It is seen that as compared to
the $TH$ modes
the curves for the hybrid modes
are much less sensitive
to variations in $n_z$.
Interestingly,
for such modes, similar to the case
of
$TH_{01}$ and $TE_{01}$ modes,
the waveguide regime may occur
in the porous films, where
$2\le k_0R\le 6$.

When the electric field is applied,
the effective refractive indices of
the LC
(in a typical situation $n_t$ decreases, whereas $n_z$ increases)
and the waveguide regime in the porous film can be suppressed.
This is obvious if the refractive index $n_t$ ,
which approaches $n_o<n_c$ as the field increases,
becomes less than $n_c$, so that
the condition of total internal reflection appears to be violated.
Figure~\ref{fig:nz-172}  shows
the dispersion curves for the case when the configuration distorted by
the field is close to axial with $n_z=n_e$, but $n_t$ is still higher than
$n_c$.
It is interesting that under these conditions, for a film with
$2\le k_0R\le 6$,
the conditions for the waveguide propagation regime are satisfied
only for the hybrid mode with $m=1$.

Radial distributions for the components
of the electric and magnetic fields
are shown
in Figs.~\ref{fig:HybE}
and~\ref{fig:HybH},
respectively.
These figures, in particular,
demonstrate that
the degree of the field
localization
inside the pore
decreases
as the director configuration
approaches the axial structure.

\section{Experiment}
\label{sec:experiment}

\subsection{Experimental procedure}
\label{subsec:procedure}

In our experiments, we have used two types of PET films
that are characterized by both micron-sized
(the diameter is $d=5$~\mum) and submicron-sized ($d= 0.6$~\mum) pores
without any surfactant treatment.
The PET films of thickness $h= 23$~\mum\ were
obtained by heavy ion bombarding with subsequent chemical etching.
Axes of  open-end pores were normal to the surface of the films. The PET films were rinsed two times with ethanol, blown, and then vacuumed to remove
a solvent. The films were filled with nematic liquid crystal (4-n-pentyl-4'-cyanobiphenyl, 5CB) in isotropic phase and then slowly cooled down to room temperature. Afterwards, the films were wiped to remove parasitic LC layers on their surfaces.

Two different cells and experimental set-ups were used for the optical
study of light propagation through the PET films filled with 5CB. In
the case of the film with submicron pores, we measured the integral
intensity of He-Ne  laser (the wavelength is $\lambda =0.633$~\mum)
beam passed through the PET film.
The linearly polarized beam was normally incident on the PET film
sandwiched between two glass substrates with
ITO electrodes and placed on an optical bench.
The intensity of light was measured at
fixed angles $\beta$ between the optical axis of the film and the
polarization plane of the beam using the scheme
without an analyzer to
avoid the birefringence effects.

In the films with micron pores, we studied the optical
textures of 5CB within the individual micron-sized pores of the PET
film. In this case, the film was stacked between two glass plates. The
lower glass plate contained ITO layer with inter electrode gap of
the width g= 50~\mum\ needed to arrange in-plane electric field, applied to
the sample. Polarized optical microscopy (POM) with 1000х
magnification was used to study optical textures of 5CB in both
crossed and parallel polarizers for
the LC cell differently oriented with respect to them.

\subsection{Experimental results and discussion}
\label{subsec:expresults}

\begin{figure*}[!tbh]
\centering
\resizebox{90mm}{!}{\includegraphics*{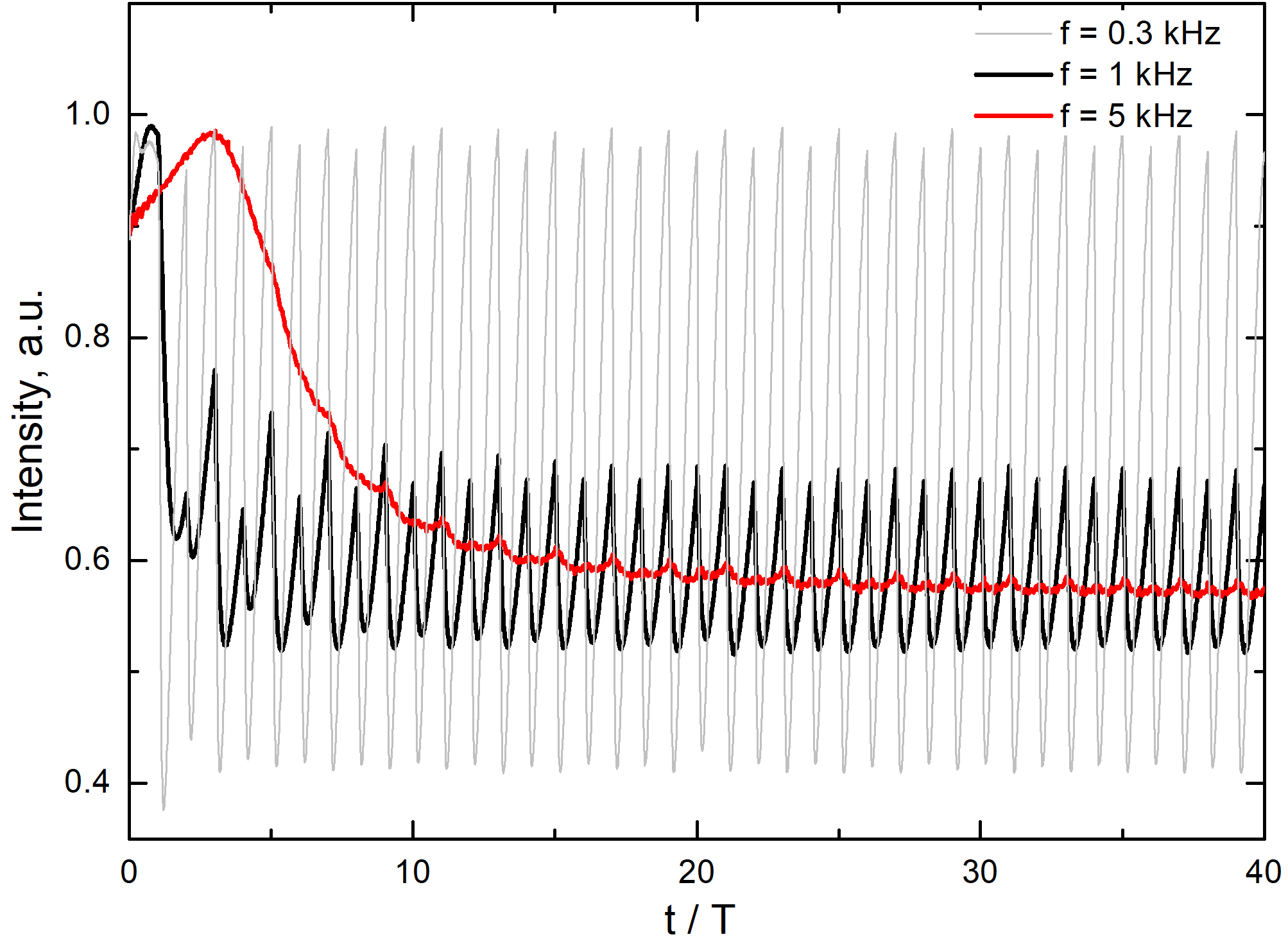}}
\caption{%
The dependence of a light intensity $I$ passed through the PET-5CB film as a function of the dimensionless time parameter $t/T$ at different
frequencies $f=0.3 \dots 5$~kHz of the applied electric field: $T=1/2f$, the voltage $U=170$~V.  
}
\label{fig:expt-1}
\end{figure*}

The electro-optical response of the sample at different frequencies of
the applied field is shown in Fig.~\ref{fig:expt-1}.
At high frequency ($f=5$~kHz),
it looks like a typical response of planar layers of LC with the positive
dielectric permittivity anisotropy in crossed polarizers, oriented at
45\dega with respect to the optical axis.
So, the dielectric mechanism of
action of electric field on LC, giving rise to overall rotation of the
director towards to the pore’s axis, is responsible for the observed
effect.

In particular, the field induced distortions of the initial
configuration results in the decrease of the effective anisotropy of
LC, introduced by the theoretical model, and disappearance of the
waveguide modes, sensitive to this parameter. It leads to decreasing
of the light intensity registered in the experiment. In the case of
strong electric fields, one can expect the breaking of
surface anchoring~\cite{Tkachenko:jopt:2008}
and formation of the axial LC configuration.
This configuration is characterized by
the ordinary refractive index which is lower than
the indexes of the polymer matrix,
so that the
waveguide regime appears to be suppressed. 
 The effective value of the electric field strength
 (of order 1 V/\mum) is comparable with
 the critical values for such
orientational transformation obtained via
the computer simulation~\cite{Tkachenko:jopt:2008}
for a nematic mixture E7 with material parameters
close to those for 5CB.

So, the breaking of the surface anchoring with
corresponding light scattering can be responsible for strong
decrease of the light intensity observed in our experiments.
It is also important to notice that such type of response is drastically
different from the low frequency response shown in
Fig.~\ref{fig:expt-1}.
This response can be explained by the additional influence of
ion's motion~\cite{Pasechnik:dd:2016}.
Nevertheless, the strong decrease of the instant values of
light intensity, registered in this case, can be also explained by the
above mentioned mechanism. 

\begin{figure*}[!tbh]
\centering
\resizebox{90mm}{!}{\includegraphics*{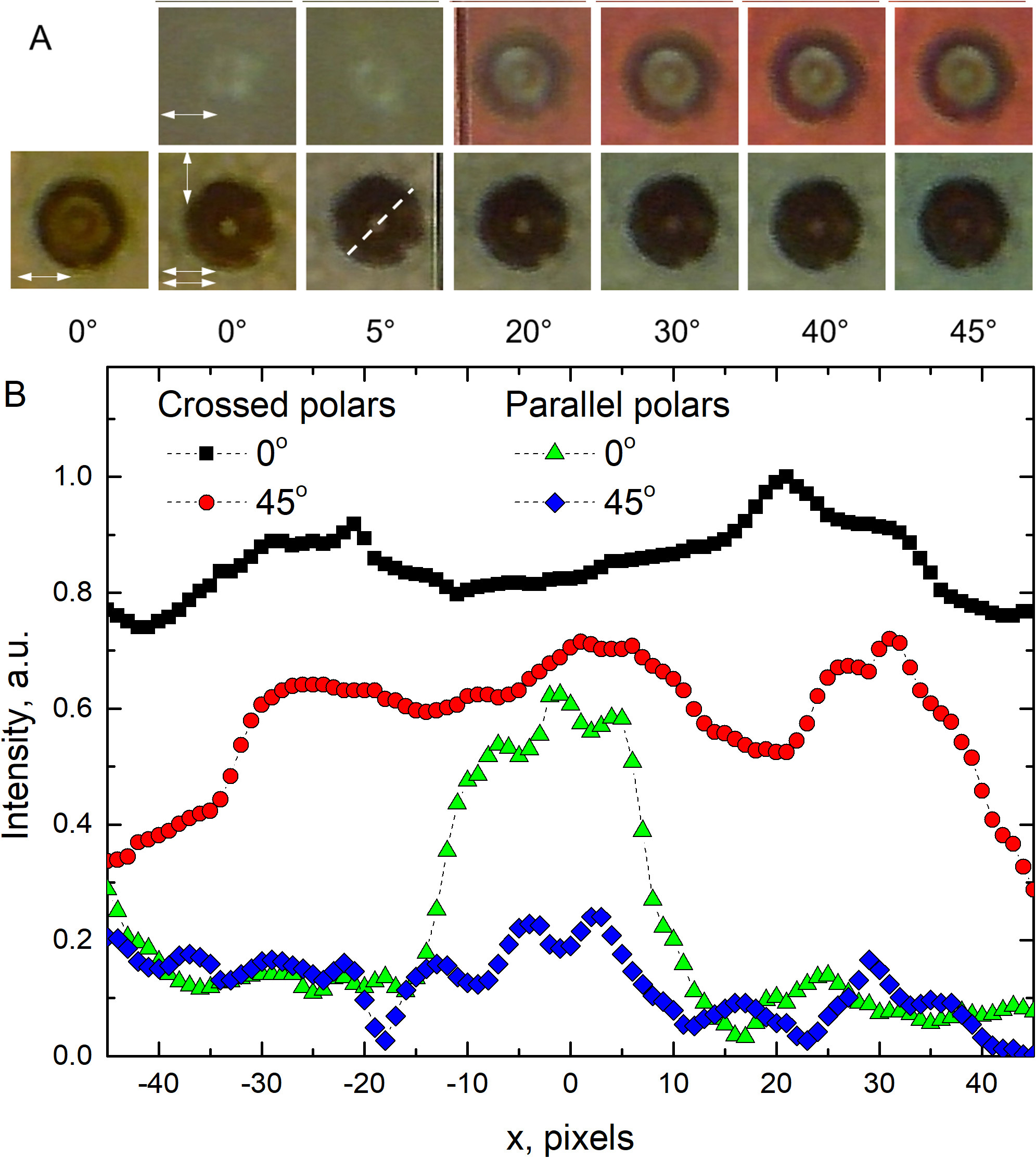}}
\caption{%
(A)~Microscopic images  of a single pore of the PET film filled with LC
and (B)~the intensity profiles  measured along the dashed line
for the film rotated by different angles
with respect to the polarizers. 
}
\label{fig:expt-2}
\end{figure*}

We also directly studied the propagation of light through the
single micrometer-sized pores ($d= 5$~\mum) of the PET film filled with LC
by analyzing the optical textures.
Figure~\ref{fig:expt-2}A shows the microscopic images of the
PET-LC film differently oriented with respect  to the crossed (parallel)
polarizers.
For the crossed polarizers,
the majority of pores  reveals
the cross-like texture indicating orientation of the director field
either along or perpendicular to the dark lines.
Such behavior may take place
as a result of different director configurations within pores.

In contrast
to the case of crossed polarizers,
a bright state appears
for the film rotated about the normal by 45 degrees.
This points to the existence of the
symmetry axis for the director configuration,
which is typical for both planar (the planar polar configuration) and
homeotropic (the axial polar configuration) anchoring. At the same time,
the dark-to-bright transition
as the film is rotated with respect to the crossed polarizers
results in the formation of the ring in the central region of the
pores.
It should
also be noted that the bright spot surrounded by the dark area within
the pores was observed in the parallel polarizers
(see the lower panel of Fig.~\ref{fig:expt-2}A).
Usually observation of the dark areas in the parallel
polarizers points to the twisting orientational LC structure.
Observation of the bright spot surrounded by the dark ring
can be explained as a result of waveguide propagation of light in the
pore.
Since the
defectless cross-like figure in crossed polarizers corresponds to the
uniform LC structure within the pores, observation of the dark area
surrounding the bright spot indicates violation of the propagation of
polarization light. In this case, the distribution of brightness
(the light intensity as we assume in this study) strongly depends on
the polarization state of light (see Fig.~\ref{fig:expt-2}B). 

\begin{figure*}[!tbh]
\centering
\resizebox{90mm}{!}{\includegraphics*{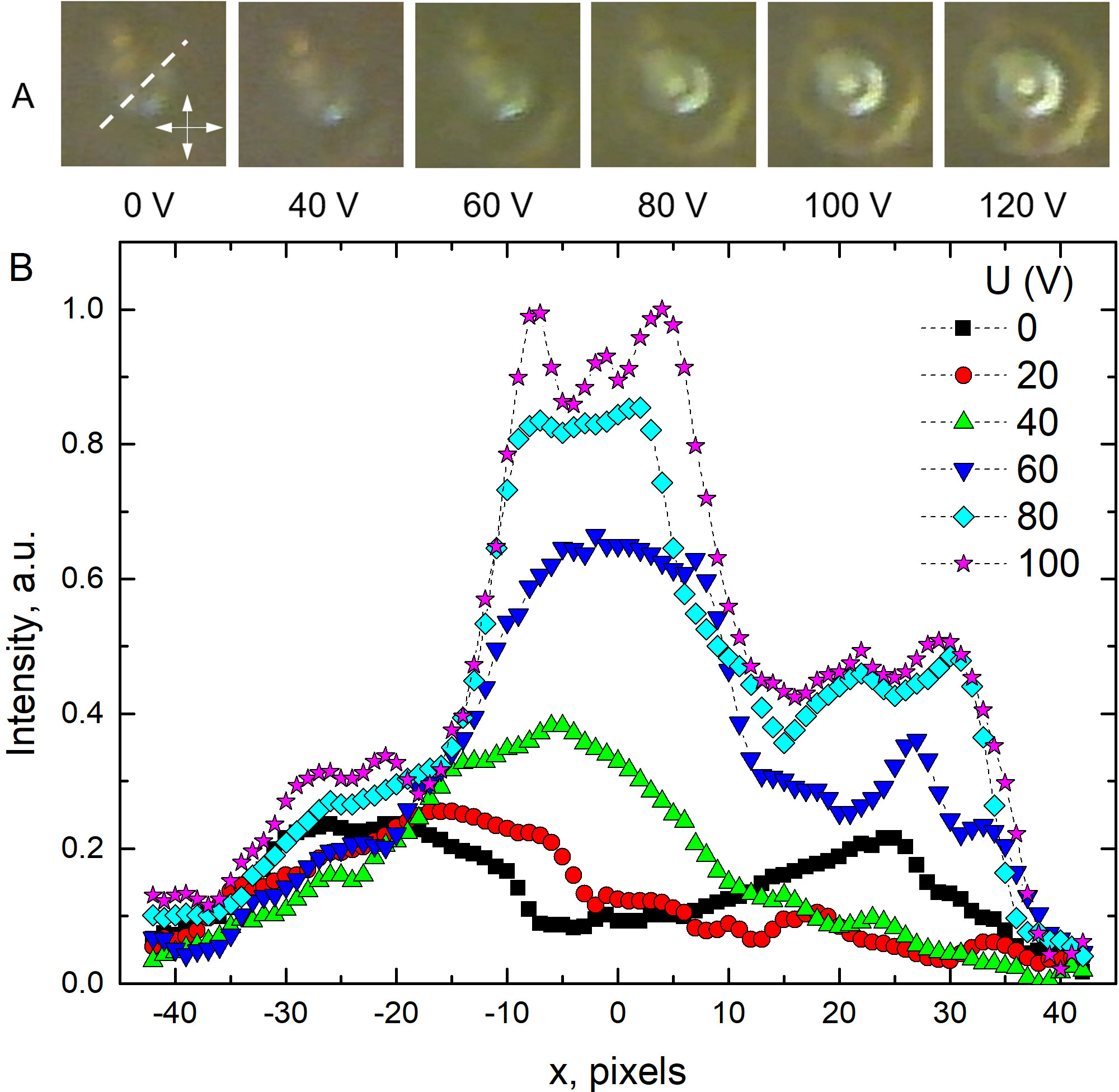}}
\caption{%
  (A)~Microscopic images  of a single pore filled with LC
  for different values of the
applied voltage (the voltage $U$ is ranged from 0 to 120~V
and $f= 3$~kHz)
and
(B)~the intensity profiles  measured along the dashed line. 
}
\label{fig:expt-3}
\end{figure*}

The dark-to-bright transition was also caused by 
the in-plane ac electric field applied to the PET-LC film
(see Fig.~\ref{fig:expt-3}A). In this case,
the electric field induced the twist torque the reorients the bulk
director of LC within the pores in the plane of the PET film.
It produces the rotation of the symmetry axis of the LC structure along
the field lines. Therefore, the symmetry axis of LC configuration twisted
from the side of the cell with inter-electrode gap to the glass
substrate without ITO that mimics the helicoid LC structure within the
pores. 

Since the electric coherence length (the length that characterizes the
thickness of a boundary layer, with the orientation stabilized by
surface in the presence of electric field) decreases with the voltage,
it explains the increase in the intensity of light passed through the
single pore (see Fig.~\ref{fig:expt-3}B).

From the above discussion,
we arrive at the conclusion that the waveguide light propagation
strongly depends on the
orientational LC configuration formed inside the pores
and can be effectively
controlled by applying either in-plane or out-of-plane electric
fields to the PET-LC film. 

\section{Conclusion}
\label{sec:conclusion}

In this paper we have theoretically studied the light propagation in a
cylindrical waveguide filled with the material characterized by the
biaxial radial optical anisotropy. The analysis of the obtained
dispersion relations for the waveguide modes showed that, when the
electric field is applied to the porous film, the waveguide
propagation regime is possible for the hybrid mode with the azimuthal
number $m=1$, until the total internal reflection condition is
violated. We have also presented some experimental results on the
waveguide regime of light propagation in polymer PET porous films with
micro-sized and submicro-sized pores filled with the nematic liquid
crystal 5CB in the presence of electric fields. In particular, we have
found that the pronounced decrease of the transmitted light intensity
induced by the strong electric field applied along the submicro-sized
pore's axis can be explained by the suppression of the propagating
waveguide modes resulting from the surface anchoring breaking and the
formation of the axial configuration. At the same time, in-plane
electric field, applied to the micro-sized pores results in essential
changes of the inhomogeneous light intensity distribution across the
pore’s cross section, which can be assigned to the electrically
induced twist deformation of the initial director configuration.

The part of our analysis was concerned with
the special case where
$\epsilon_{\rho}=\epsilon_\phi\equiv\epsilon_t$.
It relies on the assumption
that the optical properties of the pores
can be described by
the effective dielectric tensor
averaged over the LC director distribution:
$\epsilon_{ij}^{(\ind{eff})}=\epsilon_o\delta_{ij}+
(\epsilon_e-\epsilon_o)\avr{d_id_j}$
In Ref.~\cite{Tkachenko:jopt:2008}, this approximation was applied to
the escaped radial configuration
in 5CB cells with the director field of the form:
$\uvc{d}=\cos\chi(\rho)\,\uvc{e}_\rho+\sin\chi(\rho)\,\uvc{z}$
characterized by the angle $\chi(\rho)$.
In this case, the principle values of the effective dielectric tensor
are:
$\epsilon_t=\epsilon_o+(\epsilon_e-\epsilon_o)\avr{\cos^2\chi}/2$
and
$\epsilon_{zz}=\epsilon_o+(\epsilon_e-\epsilon_o)\avr{\sin^2\chi}$,
where
$\avr{\cos^2\chi}=2R^{-2}\int_0^R\cos^2\chi(\rho)\,\rho\dd\rho$.

Heuristically, validity of the above approximation
can be justified using the condition of
weakly inhomogeneous anisotropy
that can be written in the form
\begin{align}
  \label{eq:delta_n}
  k_0R\delta n<1,
\end{align}
where
$\delta n=\max|\sqrt{\bs{\epsilon}_{\ind{eff}}}-\sqrt{\bs{\epsilon}}|$
is the refractive index variation
characterizing
the difference between
the effective and local values of the refractive indices.
For the escaped radial configuration
in the 5CB cell 
without applied electric field,
it is not difficult to estimate the refractive index variation
at about $\delta n\approx 0.1$.
In deriving this estimate, we have assumed
strong homeotropic anchoring conditions
so that $\cos\chi(R)$ is close to unity.
Electric field applied across the film and
weak anchoring conditions
are the factors that may considerably reduce $\delta n$.
Our results, however, indicate that,
for micron-sized pores,
the approximation based on
the effective dielectric tensor breaks down.

Our concluding remark is that, in addition to the description of the
waveguide regime in optically anisotropic cylindrical waveguides, our
analytical results can be used to extend the class of exactly solvable
problems of the theory of light scattering. Further investigations
suggest solving the problem of light transmission for optically
anisotropic porous films that, in particular, will require a more
sophisticated analysis using the methods of the effective medium
theory.

\begin{acknowledgments} 
We acknowledge 
financial support from
the Ministry of Education and Science of the Russian Federation 
(Project No.~RFMEFI58316X0058).
\end{acknowledgments} 


\begin{thebibliography}{24}%
\makeatletter
\providecommand \@ifxundefined [1]{%
 \@ifx{#1\undefined}
}%
\providecommand \@ifnum [1]{%
 \ifnum #1\expandafter \@firstoftwo
 \else \expandafter \@secondoftwo
 \fi
}%
\providecommand \@ifx [1]{%
 \ifx #1\expandafter \@firstoftwo
 \else \expandafter \@secondoftwo
 \fi
}%
\providecommand \natexlab [1]{#1}%
\providecommand \enquote  [1]{``#1''}%
\providecommand \bibnamefont  [1]{#1}%
\providecommand \bibfnamefont [1]{#1}%
\providecommand \citenamefont [1]{#1}%
\providecommand \href@noop [0]{\@secondoftwo}%
\providecommand \href [0]{\begingroup \@sanitize@url \@href}%
\providecommand \@href[1]{\@@startlink{#1}\@@href}%
\providecommand \@@href[1]{\endgroup#1\@@endlink}%
\providecommand \@sanitize@url [0]{\catcode `\\12\catcode `\$12\catcode
  `\&12\catcode `\#12\catcode `\^12\catcode `\_12\catcode `\%12\relax}%
\providecommand \@@startlink[1]{}%
\providecommand \@@endlink[0]{}%
\providecommand \url  [0]{\begingroup\@sanitize@url \@url }%
\providecommand \@url [1]{\endgroup\@href {#1}{\urlprefix }}%
\providecommand \urlprefix  [0]{URL }%
\providecommand \Eprint [0]{\href }%
\providecommand \doibase [0]{http://dx.doi.org/}%
\providecommand \selectlanguage [0]{\@gobble}%
\providecommand \bibinfo  [0]{\@secondoftwo}%
\providecommand \bibfield  [0]{\@secondoftwo}%
\providecommand \translation [1]{[#1]}%
\providecommand \BibitemOpen [0]{}%
\providecommand \bibitemStop [0]{}%
\providecommand \bibitemNoStop [0]{.\EOS\space}%
\providecommand \EOS [0]{\spacefactor3000\relax}%
\providecommand \BibitemShut  [1]{\csname bibitem#1\endcsname}%
\let\auto@bib@innerbib\@empty
\bibitem [{\citenamefont {Corella-Madueno}\ and\ \citenamefont
  {Reyes}(2006)}]{Corella:optcomm:2006}%
  \BibitemOpen
  \bibfield  {author} {\bibinfo {author} {\bibfnamefont {A.}~\bibnamefont
  {Corella-Madueno}}\ and\ \bibinfo {author} {\bibfnamefont {J.~Adri\'an}\
  \bibnamefont {Reyes}},\ }\bibfield  {title} {{\selectlanguage
  {english}\enquote {\bibinfo {title} {Electrically controlled liquid crystal
  fiber},}\ }}\href {\doibase 10.1016/j.optcom.2006.02.035} {\bibfield
  {journal} {\bibinfo  {journal} {Optics Communications}\ }\textbf {\bibinfo
  {volume} {264}},\ \bibinfo {pages} {148--155} (\bibinfo {year}
  {2006})}\BibitemShut {NoStop}%
\bibitem [{\citenamefont {Tkachenko}\ \emph {et~al.}(2008)\citenamefont
  {Tkachenko}, \citenamefont {Dyomin}, \citenamefont {Tkachenko}, \citenamefont
  {Abbate},\ and\ \citenamefont {Sukhoivanov}}]{Tkachenko:jopt:2008}%
  \BibitemOpen
  \bibfield  {author} {\bibinfo {author} {\bibfnamefont {V.}~\bibnamefont
  {Tkachenko}}, \bibinfo {author} {\bibfnamefont {A.~A.}\ \bibnamefont
  {Dyomin}}, \bibinfo {author} {\bibfnamefont {G.~V.}\ \bibnamefont
  {Tkachenko}}, \bibinfo {author} {\bibfnamefont {G.}~\bibnamefont {Abbate}}, \
  and\ \bibinfo {author} {\bibfnamefont {I.~A.}\ \bibnamefont {Sukhoivanov}},\
  }\bibfield  {title} {{\selectlanguage {english}\enquote {\bibinfo {title}
  {Electrical reorientation of liquid crystal molecules inside cylindrical
  pores for photonic device applications},}\ }}\href {\doibase
  10.1088/1464-4258/10/5/055301} {\bibfield  {journal} {\bibinfo  {journal} {J.
  Opt. A: Pure Appl. Opt. 10 (2008) 055301 (6pp)}\ }\textbf {\bibinfo {volume}
  {10}},\ \bibinfo {pages} {055301} (\bibinfo {year} {2008})}\BibitemShut
  {NoStop}%
\bibitem [{\citenamefont {Wahle}\ and\ \citenamefont
  {Kitzerow}(2014)}]{Wahle:optexp:2014}%
  \BibitemOpen
  \bibfield  {author} {\bibinfo {author} {\bibfnamefont {M.}~\bibnamefont
  {Wahle}}\ and\ \bibinfo {author} {\bibfnamefont {H.-S.}\ \bibnamefont
  {Kitzerow}},\ }\bibfield  {title} {{\selectlanguage {english}\enquote
  {\bibinfo {title} {Liquid crystal assisted optical fibres},}\ }}\href
  {\doibase 10.1364/OE.22.000262} {\bibfield  {journal} {\bibinfo  {journal}
  {Opt. Express}\ }\textbf {\bibinfo {volume} {22}},\ \bibinfo {pages}
  {262--273} (\bibinfo {year} {2014})}\BibitemShut {NoStop}%
\bibitem [{\citenamefont {Markos}\ \emph {et~al.}(2017)\citenamefont {Markos},
  \citenamefont {Travers}, \citenamefont {Abdolvand}, \citenamefont
  {Eggleton},\ and\ \citenamefont {Bang}}]{Marcos:rmp:2017}%
  \BibitemOpen
  \bibfield  {author} {\bibinfo {author} {\bibfnamefont {Christos}\
  \bibnamefont {Markos}}, \bibinfo {author} {\bibfnamefont {John~C.}\
  \bibnamefont {Travers}}, \bibinfo {author} {\bibfnamefont {Amir}\
  \bibnamefont {Abdolvand}}, \bibinfo {author} {\bibfnamefont {Benjamin~J.}\
  \bibnamefont {Eggleton}}, \ and\ \bibinfo {author} {\bibfnamefont {Ole}\
  \bibnamefont {Bang}},\ }\bibfield  {title} {{\selectlanguage
  {english}\enquote {\bibinfo {title} {Hybrid photonic-crystal fiber},}\
  }}\href {\doibase 10.1103/RevModPhys.89.045003} {\bibfield  {journal}
  {\bibinfo  {journal} {Rev. Mod. Phys.}\ }\textbf {\bibinfo {volume} {89}},\
  \bibinfo {pages} {045003} (\bibinfo {year} {2017})}\BibitemShut {NoStop}%
\bibitem [{\citenamefont {Beeckman}\ \emph {et~al.}(2011)\citenamefont
  {Beeckman}, \citenamefont {Neyts},\ and\ \citenamefont
  {Vanbrabant}}]{Beeckman:opteng:2011}%
  \BibitemOpen
  \bibfield  {author} {\bibinfo {author} {\bibfnamefont {Jeroen}\ \bibnamefont
  {Beeckman}}, \bibinfo {author} {\bibfnamefont {Kristiaan}\ \bibnamefont
  {Neyts}}, \ and\ \bibinfo {author} {\bibfnamefont {Pieter J.~M.}\
  \bibnamefont {Vanbrabant}},\ }\bibfield  {title} {{\selectlanguage
  {english}\enquote {\bibinfo {title} {Liquid-crystal photonic applications},}\
  }}\href {\doibase 10.1117/1.3565046} {\bibfield  {journal} {\bibinfo
  {journal} {Optical Engineering}\ }\textbf {\bibinfo {volume} {50}},\ \bibinfo
  {pages} {081202} (\bibinfo {year} {2011})}\BibitemShut {NoStop}%
\bibitem [{\citenamefont {Pasechnik}\ \emph {et~al.}(2009)\citenamefont
  {Pasechnik}, \citenamefont {Chigrinov},\ and\ \citenamefont
  {Shmeliova}}]{Pasechnik:bk:2009}%
  \BibitemOpen
  \bibfield  {author} {\bibinfo {author} {\bibfnamefont {S.~V.}\ \bibnamefont
  {Pasechnik}}, \bibinfo {author} {\bibfnamefont {V.~G.}\ \bibnamefont
  {Chigrinov}}, \ and\ \bibinfo {author} {\bibfnamefont {D.~V.}\ \bibnamefont
  {Shmeliova}},\ }\href@noop {} {{\selectlanguage {english}\emph {\bibinfo
  {title} {Liquid Crystals: Viscous and Elastic Properties}}}}\ (\bibinfo
  {publisher} {Wiley-VCH},\ \bibinfo {address} {Berlin},\ \bibinfo {year}
  {2009})\ p.\ \bibinfo {pages} {424}\BibitemShut {NoStop}%
\bibitem [{\citenamefont {Carlton}\ \emph {et~al.}(2013)\citenamefont
  {Carlton}, \citenamefont {Hunter}, \citenamefont {Miller}, \citenamefont
  {Abbasi}, \citenamefont {Mushenheim}, \citenamefont {Tan},\ and\
  \citenamefont {Abbott}}]{Carlton:lcrev:2013}%
  \BibitemOpen
  \bibfield  {author} {\bibinfo {author} {\bibfnamefont {Rebecca~J.}\
  \bibnamefont {Carlton}}, \bibinfo {author} {\bibfnamefont {Jacob~T.}\
  \bibnamefont {Hunter}}, \bibinfo {author} {\bibfnamefont {Daniel~S.}\
  \bibnamefont {Miller}}, \bibinfo {author} {\bibfnamefont {Reza}\ \bibnamefont
  {Abbasi}}, \bibinfo {author} {\bibfnamefont {Peter~C.}\ \bibnamefont
  {Mushenheim}}, \bibinfo {author} {\bibfnamefont {Lie~Na}\ \bibnamefont
  {Tan}}, \ and\ \bibinfo {author} {\bibfnamefont {Nicholas~L.}\ \bibnamefont
  {Abbott}},\ }\bibfield  {title} {{\selectlanguage {english}\enquote {\bibinfo
  {title} {Chemical and biological sensing using liquid crystals},}\ }}\href
  {\doibase 10.1080/21680396.2013.769310} {\bibfield  {journal} {\bibinfo
  {journal} {Liquid Crystals Reviews}\ }\textbf {\bibinfo {volume} {1}},\
  \bibinfo {pages} {29--51} (\bibinfo {year} {2013})}\BibitemShut {NoStop}%
\bibitem [{\citenamefont {Pan}\ \emph {et~al.}(2004)\citenamefont {Pan},
  \citenamefont {Chen}, \citenamefont {Tsai},\ and\ \citenamefont
  {Pan}}]{Pan:mclc:2004}%
  \BibitemOpen
  \bibfield  {author} {\bibinfo {author} {\bibfnamefont {Ru-Pin}\ \bibnamefont
  {Pan}}, \bibinfo {author} {\bibfnamefont {Chao-Yuan}\ \bibnamefont {Chen}},
  \bibinfo {author} {\bibfnamefont {Tsong-Ru}\ \bibnamefont {Tsai}}, \ and\
  \bibinfo {author} {\bibfnamefont {Ci-Ling}\ \bibnamefont {Pan}},\ }\bibfield
  {title} {{\selectlanguage {english}\enquote {\bibinfo {title} {Terahertz
  phase shifter with nematic liquid crystal in a magnetic field},}\ }}\href
  {\doibase 10.1080/15421400490501752} {\bibfield  {journal} {\bibinfo
  {journal} {Molecular Crystals and Liquid Crystals}\ }\textbf {\bibinfo
  {volume} {421}},\ \bibinfo {pages} {157--164} (\bibinfo {year}
  {2004})}\BibitemShut {NoStop}%
\bibitem [{\citenamefont {Larsen}\ \emph {et~al.}(2003)\citenamefont {Larsen},
  \citenamefont {Bjarklev}, \citenamefont {Hermann},\ and\ \citenamefont
  {Broeng}}]{Larsen:optexp:2003}%
  \BibitemOpen
  \bibfield  {author} {\bibinfo {author} {\bibfnamefont {Thomas~Tanggaard}\
  \bibnamefont {Larsen}}, \bibinfo {author} {\bibfnamefont {Anders}\
  \bibnamefont {Bjarklev}}, \bibinfo {author} {\bibfnamefont {David~Sparre}\
  \bibnamefont {Hermann}}, \ and\ \bibinfo {author} {\bibfnamefont {Jes}\
  \bibnamefont {Broeng}},\ }\bibfield  {title} {{\selectlanguage
  {english}\enquote {\bibinfo {title} {Optical devices based on liquid crystal
  photonic bandgap fibres},}\ }}\href {\doibase 10.1364/OE.11.002589}
  {\bibfield  {journal} {\bibinfo  {journal} {Opt. Express}\ }\textbf {\bibinfo
  {volume} {11}},\ \bibinfo {pages} {2589--2596} (\bibinfo {year}
  {2003})}\BibitemShut {NoStop}%
\bibitem [{\citenamefont {Semerenko}\ \emph {et~al.}(2010)\citenamefont
  {Semerenko}, \citenamefont {Shmeliova}, \citenamefont {Pasechnik},
  \citenamefont {Murauskii}, \citenamefont {Tsvetkov},\ and\ \citenamefont
  {Chigrinov}}]{Semerenko:ol:2010}%
  \BibitemOpen
  \bibfield  {author} {\bibinfo {author} {\bibfnamefont {Denis}\ \bibnamefont
  {Semerenko}}, \bibinfo {author} {\bibfnamefont {Dina}\ \bibnamefont
  {Shmeliova}}, \bibinfo {author} {\bibfnamefont {Sergey}\ \bibnamefont
  {Pasechnik}}, \bibinfo {author} {\bibfnamefont {Anatolii}\ \bibnamefont
  {Murauskii}}, \bibinfo {author} {\bibfnamefont {Valentin}\ \bibnamefont
  {Tsvetkov}}, \ and\ \bibinfo {author} {\bibfnamefont {Vladimir}\ \bibnamefont
  {Chigrinov}},\ }\bibfield  {title} {{\selectlanguage {english}\enquote
  {\bibinfo {title} {Optically controlled transmission of porous polyethylene
  terephthalate films filled with nematic liquid crystal},}\ }}\href {\doibase
  10.1364/OL.35.002155} {\bibfield  {journal} {\bibinfo  {journal} {Opt.
  Lett.}\ }\textbf {\bibinfo {volume} {35}},\ \bibinfo {pages} {2155--2157}
  (\bibinfo {year} {2010})}\BibitemShut {NoStop}%
\bibitem [{\citenamefont {Chopik}\ \emph {et~al.}(2014)\citenamefont {Chopik},
  \citenamefont {Pasechnik}, \citenamefont {Semerenko}, \citenamefont
  {Shmeliova}, \citenamefont {Dubtsov}, \citenamefont {Srivastava},\ and\
  \citenamefont {Chigrinov}}]{Chopik:optlett:2014}%
  \BibitemOpen
  \bibfield  {author} {\bibinfo {author} {\bibfnamefont {A.}~\bibnamefont
  {Chopik}}, \bibinfo {author} {\bibfnamefont {S.}~\bibnamefont {Pasechnik}},
  \bibinfo {author} {\bibfnamefont {D.}~\bibnamefont {Semerenko}}, \bibinfo
  {author} {\bibfnamefont {D.}~\bibnamefont {Shmeliova}}, \bibinfo {author}
  {\bibfnamefont {A.}~\bibnamefont {Dubtsov}}, \bibinfo {author} {\bibfnamefont
  {A.~K.}\ \bibnamefont {Srivastava}}, \ and\ \bibinfo {author} {\bibfnamefont
  {V.}~\bibnamefont {Chigrinov}},\ }\bibfield  {title} {{\selectlanguage
  {english}\enquote {\bibinfo {title} {Electro-optical effects in porous {PET}
  films filled with liquid crystal: new possibilities for fiber optics and
  {THZ} applications},}\ }}\href {\doibase 10.1364/OL.39.001453} {\bibfield
  {journal} {\bibinfo  {journal} {Optics Letters}\ }\textbf {\bibinfo {volume}
  {39}},\ \bibinfo {pages} {1453--1456} (\bibinfo {year} {2014})}\BibitemShut
  {NoStop}%
\bibitem [{\citenamefont {Pasechnik}\ \emph {et~al.}(2016)\citenamefont
  {Pasechnik}, \citenamefont {Shmeliova}, \citenamefont {Chopik}, \citenamefont
  {Semerenko}, \citenamefont {Kharlamov},\ and\ \citenamefont
  {Dubtsov}}]{Pasechnik:dd:2016}%
  \BibitemOpen
  \bibfield  {author} {\bibinfo {author} {\bibfnamefont {S.~V.}\ \bibnamefont
  {Pasechnik}}, \bibinfo {author} {\bibfnamefont {D.~V.}\ \bibnamefont
  {Shmeliova}}, \bibinfo {author} {\bibfnamefont {A.~P.}\ \bibnamefont
  {Chopik}}, \bibinfo {author} {\bibfnamefont {D.~A.}\ \bibnamefont
  {Semerenko}}, \bibinfo {author} {\bibfnamefont {S.~S.}\ \bibnamefont
  {Kharlamov}}, \ and\ \bibinfo {author} {\bibfnamefont {A.~V.}\ \bibnamefont
  {Dubtsov}},\ }\bibfield  {title} {{\selectlanguage {english}\enquote
  {\bibinfo {title} {{Electrically controlled porous polymer films filled with
  liquid crystals: New possibilities for photonics and THz applications}},}\
  }}in\ \href {\doibase 10.1109/DD.2016.7756864} {{\selectlanguage
  {english}\emph {\bibinfo {booktitle} {2016 Days on Diffraction (DD)}}}}\
  (\bibinfo {year} {2016})\ pp.\ \bibinfo {pages} {314--318}\BibitemShut
  {NoStop}%
\bibitem [{\citenamefont {Lin}\ and\ \citenamefont
  {Palffy-Muhoray}(1992)}]{Lin:ol:1992}%
  \BibitemOpen
  \bibfield  {author} {\bibinfo {author} {\bibfnamefont {H.}~\bibnamefont
  {Lin}}\ and\ \bibinfo {author} {\bibfnamefont {P.}~\bibnamefont
  {Palffy-Muhoray}},\ }\bibfield  {title} {{\selectlanguage {english}\enquote
  {\bibinfo {title} {{TE} and {TM} modes in a cylindrical liquid-crystal
  waveguide},}\ }}\href {\doibase 10.1364/OL.17.000722} {\bibfield  {journal}
  {\bibinfo  {journal} {Opt. Lett.}\ }\textbf {\bibinfo {volume} {17}},\
  \bibinfo {pages} {722--724} (\bibinfo {year} {1992})}\BibitemShut {NoStop}%
\bibitem [{\citenamefont {Reyes}\ and\ \citenamefont
  {Rodr\'{\i}guez}(2002)}]{Reyes:pre:2002}%
  \BibitemOpen
  \bibfield  {author} {\bibinfo {author} {\bibfnamefont {J.~A.}\ \bibnamefont
  {Reyes}}\ and\ \bibinfo {author} {\bibfnamefont {R.~F.}\ \bibnamefont
  {Rodr\'{\i}guez}},\ }\bibfield  {title} {{\selectlanguage {english}\enquote
  {\bibinfo {title} {Flow dissipation effects in a nonlinear nematic fiber},}\
  }}\href {\doibase 10.1103/PhysRevE.65.051701} {\bibfield  {journal} {\bibinfo
   {journal} {Phys. Rev. E}\ }\textbf {\bibinfo {volume} {65}},\ \bibinfo
  {pages} {051701} (\bibinfo {year} {2002})}\BibitemShut {NoStop}%
\bibitem [{\citenamefont {Rodr\'{\i}guez}\ \emph {et~al.}(2003)\citenamefont
  {Rodr\'{\i}guez}, \citenamefont {Reyes}, \citenamefont {Espinosa-Cer\'on},
  \citenamefont {Fujioka},\ and\ \citenamefont {Malomed}}]{Rodrigues:pre:2003}%
  \BibitemOpen
  \bibfield  {author} {\bibinfo {author} {\bibfnamefont {R.~F.}\ \bibnamefont
  {Rodr\'{\i}guez}}, \bibinfo {author} {\bibfnamefont {J.~A.}\ \bibnamefont
  {Reyes}}, \bibinfo {author} {\bibfnamefont {A.}~\bibnamefont
  {Espinosa-Cer\'on}}, \bibinfo {author} {\bibfnamefont {J.}~\bibnamefont
  {Fujioka}}, \ and\ \bibinfo {author} {\bibfnamefont {B.~A.}\ \bibnamefont
  {Malomed}},\ }\bibfield  {title} {{\selectlanguage {english}\enquote
  {\bibinfo {title} {Standard and embedded solitons in nematic optical
  fibers},}\ }}\href {\doibase 10.1103/PhysRevE.68.036606} {\bibfield
  {journal} {\bibinfo  {journal} {Phys. Rev. E}\ }\textbf {\bibinfo {volume}
  {68}},\ \bibinfo {pages} {036606} (\bibinfo {year} {2003})}\BibitemShut
  {NoStop}%
\bibitem [{\citenamefont {Corella-Madueno}\ and\ \citenamefont
  {Reyes}(2008)}]{Corella:physb:2008}%
  \BibitemOpen
  \bibfield  {author} {\bibinfo {author} {\bibfnamefont {A.}~\bibnamefont
  {Corella-Madueno}}\ and\ \bibinfo {author} {\bibfnamefont {J.~Adri\'an}\
  \bibnamefont {Reyes}},\ }\bibfield  {title} {{\selectlanguage
  {english}\enquote {\bibinfo {title} {Hydrodynamically controlled optical
  propagation in a nematic fiber},}\ }}\href {\doibase
  10.1016/j.physb.2007.10.258} {\bibfield  {journal} {\bibinfo  {journal}
  {Physica B: Condensed Matter}\ }\textbf {\bibinfo {volume} {403}},\ \bibinfo
  {pages} {1949--1955} (\bibinfo {year} {2008})}\BibitemShut {NoStop}%
\bibitem [{\citenamefont {Allender}\ \emph {et~al.}(1991)\citenamefont
  {Allender}, \citenamefont {Crawford},\ and\ \citenamefont
  {Doane}}]{Alld:prl:1991}%
  \BibitemOpen
  \bibfield  {author} {\bibinfo {author} {\bibfnamefont {D.~W.}\ \bibnamefont
  {Allender}}, \bibinfo {author} {\bibfnamefont {G.~P.}\ \bibnamefont
  {Crawford}}, \ and\ \bibinfo {author} {\bibfnamefont {J.~W.}\ \bibnamefont
  {Doane}},\ }\bibfield  {title} {{\selectlanguage {english}\enquote {\bibinfo
  {title} {Determination of the liquid-crystal surface elastic constant
  {$K_{24}$}},}\ }}\href@noop {} {\bibfield  {journal} {\bibinfo  {journal}
  {Phys. Rev. Lett.}\ }\textbf {\bibinfo {volume} {67}},\ \bibinfo {pages}
  {1442--1445} (\bibinfo {year} {1991})}\BibitemShut {NoStop}%
\bibitem [{\citenamefont {Crawford}\ \emph {et~al.}(1993)\citenamefont
  {Crawford}, \citenamefont {Ondris-Crawford}, \citenamefont {\v{Z}umer},\ and\
  \citenamefont {Doane}}]{Craw:prl2:1993}%
  \BibitemOpen
  \bibfield  {author} {\bibinfo {author} {\bibfnamefont {G.~P.}\ \bibnamefont
  {Crawford}}, \bibinfo {author} {\bibfnamefont {R.~J.}\ \bibnamefont
  {Ondris-Crawford}}, \bibinfo {author} {\bibfnamefont {S.}~\bibnamefont
  {\v{Z}umer}}, \ and\ \bibinfo {author} {\bibfnamefont {J.~W.}\ \bibnamefont
  {Doane}},\ }\bibfield  {title} {{\selectlanguage {english}\enquote {\bibinfo
  {title} {Anchoring and orientational wetting transitions of confined liquid
  crystals},}\ }}\href@noop {} {\bibfield  {journal} {\bibinfo  {journal}
  {Phys. Rev. Lett.}\ }\textbf {\bibinfo {volume} {70}},\ \bibinfo {pages}
  {1838--1841} (\bibinfo {year} {1993})}\BibitemShut {NoStop}%
\bibitem [{\citenamefont {Kralj}\ and\ \citenamefont
  {Zumer}(1993)}]{Kralj:lc:1993}%
  \BibitemOpen
  \bibfield  {author} {\bibinfo {author} {\bibfnamefont {S.}~\bibnamefont
  {Kralj}}\ and\ \bibinfo {author} {\bibfnamefont {S.}~\bibnamefont {Zumer}},\
  }\bibfield  {title} {{\selectlanguage {english}\enquote {\bibinfo {title}
  {The stability diagram of a nematic liquid crystal confined to a cylindrical
  cavity},}\ }}\href {\doibase 10.1080/02678299308036471} {\bibfield  {journal}
  {\bibinfo  {journal} {Liquid Crystals}\ }\textbf {\bibinfo {volume} {15}},\
  \bibinfo {pages} {521--527} (\bibinfo {year} {1993})}\BibitemShut {NoStop}%
\bibitem [{\citenamefont {Kiselev}\ and\ \citenamefont
  {Reshetnyak}(1995{\natexlab{a}})}]{Kiselev:jetp:1995}%
  \BibitemOpen
  \bibfield  {author} {\bibinfo {author} {\bibfnamefont {A.~D.}\ \bibnamefont
  {Kiselev}}\ and\ \bibinfo {author} {\bibfnamefont {V.~Yu.}\ \bibnamefont
  {Reshetnyak}},\ }\bibfield  {title} {{\selectlanguage {english}\enquote
  {\bibinfo {title} {Stability of the axial director configuration in nematic
  liquid crystals confined in a cylindrical cavity and the surface-like elastic
  constant problem},}\ }}\href@noop {} {\bibfield  {journal} {\bibinfo
  {journal} {JETP}\ }\textbf {\bibinfo {volume} {107}},\ \bibinfo {pages}
  {1552--1562} (\bibinfo {year} {1995}{\natexlab{a}})}\BibitemShut {NoStop}%
\bibitem [{\citenamefont {Kiselev}\ and\ \citenamefont
  {Reshetnyak}(1995{\natexlab{b}})}]{Kiselev:mclc:1995}%
  \BibitemOpen
  \bibfield  {author} {\bibinfo {author} {\bibfnamefont {A.~D.}\ \bibnamefont
  {Kiselev}}\ and\ \bibinfo {author} {\bibfnamefont {V.~Yu.}\ \bibnamefont
  {Reshetnyak}},\ }\bibfield  {title} {{\selectlanguage {english}\enquote
  {\bibinfo {title} {Surface elastic constant problems for {NLC} confined to
  cylindrical cavity: {S}tability of axial configuration},}\ }}\href@noop {}
  {\bibfield  {journal} {\bibinfo  {journal} {Mol. Cryst. Liq. Cryst.}\
  }\textbf {\bibinfo {volume} {265}},\ \bibinfo {pages} {527--540} (\bibinfo
  {year} {1995}{\natexlab{b}})}\BibitemShut {NoStop}%
\bibitem [{\citenamefont {Jin}\ \emph {et~al.}(2010)\citenamefont {Jin},
  \citenamefont {Gao},\ and\ \citenamefont {Gao}}]{Gao:jpier:2010}%
  \BibitemOpen
  \bibfield  {author} {\bibinfo {author} {\bibfnamefont {Y.~W.}\ \bibnamefont
  {Jin}}, \bibinfo {author} {\bibfnamefont {D.~L.}\ \bibnamefont {Gao}}, \ and\
  \bibinfo {author} {\bibfnamefont {L.}~\bibnamefont {Gao}},\ }\bibfield
  {title} {{\selectlanguage {english}\enquote {\bibinfo {title} {Plasmonic
  resonant light scattering by a cylinder with radial anisotropy},}\ }}\href
  {\doibase 10.2528/PIER10060601} {\bibfield  {journal} {\bibinfo  {journal}
  {Progress In Electromagnetics Research}\ }\textbf {\bibinfo {volume} {106}},\
  \bibinfo {pages} {335--347} (\bibinfo {year} {2010})}\BibitemShut {NoStop}%
\bibitem [{\citenamefont {Chen}\ and\ \citenamefont
  {Gao}(2012)}]{Chen:pra:2012}%
  \BibitemOpen
  \bibfield  {author} {\bibinfo {author} {\bibfnamefont {H.~L.}\ \bibnamefont
  {Chen}}\ and\ \bibinfo {author} {\bibfnamefont {L.}~\bibnamefont {Gao}},\
  }\bibfield  {title} {{\selectlanguage {english}\enquote {\bibinfo {title}
  {Anomalous electromagnetic scattering from radially anisotropic nanowires},}\
  }}\href {\doibase 10.1103/PhysRevA.86.033825} {\bibfield  {journal} {\bibinfo
   {journal} {Phys. Rev. A}\ }\textbf {\bibinfo {volume} {86}},\ \bibinfo
  {pages} {033825} (\bibinfo {year} {2012})}\BibitemShut {NoStop}%
\bibitem [{\citenamefont {Abramowitz}\ and\ \citenamefont
  {Stegun}(1972)}]{Abr}%
  \BibitemOpen
  \bibinfo {editor} {\bibfnamefont {M.}~\bibnamefont {Abramowitz}}\ and\
  \bibinfo {editor} {\bibfnamefont {I.~A.}\ \bibnamefont {Stegun}},\ eds.,\
  \href@noop {} {{\selectlanguage {english}\emph {\bibinfo {title} {Handbook of
  Mathematical Functions}}}}\ (\bibinfo  {publisher} {Dover},\ \bibinfo
  {address} {New York},\ \bibinfo {year} {1972})\BibitemShut {NoStop}%
\end{thebibliography}

%

\end{document}